\documentclass[preprint,12pt]{article}

\usepackage{amssymb}

\title{Radiative corrections of the order $\mathcal{O}(\alpha^3L^3)$ to unpolarized muon decay spectrum}

\author{A.B. Arbuzov$^{a,b}$\footnote{arbuzov@theor.jinr.ru},
U.E. Voznaya$^{a,b}$\footnote{voznaya@theor.jinr.ru}}

\begin{document}
	
\sloppypar	

\maketitle

\begin{center}
{\it
$^a$Joint Institute for Nuclear Research,
            Joliot-Curie str. 6, Dubna, 141980, Moscow region, Russia \\
$^b$Dubna state university, Universitetskaya str. 19, Dubna, 141980, Moscow region, Russia
}
\end{center}

\begin{abstract}
Calculation of higher-order radiative corrections to unpolarized muon decay spectrum is discussed. Results for the orders $\mathcal{O}(\alpha^2 L)$, $\mathcal{O}(\alpha^3 L^3)$ and $\mathcal{O}(\alpha^3 L^2)$ are presented.

PACS: 12.20.-m, 13.40.-f 
\end{abstract}

\section*{Introduction}
The process of muon decay is almost a pure weak-interaction process with small QED, QCD, 
and possibly new physics additions. In high-precision experiments with muons even small 
deviations from the Standard Model (SM) pointing to new physics can be seen.
For such experiments, accurate theoretical predictions within the SM are needed. 
In this work corrections to unpolarized muon decay spectrum up to the order 
$\mathcal{O}(\alpha^3 L^2)$ are presented.

The article is based on the talk given by Uliana Voznaya at the XXVII International Scientific 
Conference of Young Scientists and Specialists (AYSS-2023) in JINR, Dubna, Russia.

\section*{Parton distribution function approach}

To compute higher order corrections to the muon decay spectrum we use here 
the QED structure (or parton distribution) function approach~\cite{Kuraev:1985hb}.
In this approach expansion is made not only in the powers of the coupling constant but also in the ones of the large logarithm $L=\ln(\mu^2_F/\mu^2_R)$
where $\mu_F$ is the factorization scale and $\mu_R$ is the renormalization scale. For muon decay 
we take usually $\mu_F=m_\mu$ and $\mu_R=m_e$, so  $L = \ln(m_{\mu}^2/m_e^2) \approx 10.66$.

In the QED PDF approach we use universal process independent parton distribution and fragmentation functions. To calculate radiative corrections to a particular process we have to make a convolution of them with the Wilson coefficients containing information about the process. For unpolarized muon decay, these coeficient functions are~\cite{Kinoshita:1958ru}
\begin{eqnarray}
	&& f_e^{(0)} (z)= z^2 (3 - 2z), \quad f_\gamma^{(0)} (z)= 0, \nonumber \\
	&& f_e^{(1)} (z) = 2 z^2 (2 z-3) \bigl( 4 \zeta(2) - 4 \mathrm{Li}_2 (z) + 2  \ln z^2 
	- 3  \ln z  \ln (1-z)  \nonumber
	\\ \nonumber
	&&   -  \ln (1-z)^2 \bigr) + \left( \frac{5}{3} - 2 z - 13 z^2 + \frac{34}{3} z^3 \right)  \ln (1-z)
	\\
	&&   + \left( \frac{5}{3} + 4 z - 2 z^2 - 6 z^3 \right)  \ln z  + \frac{5}{6} - \frac{23}{3} z -\frac{3}{2} z^2 + \frac{7}{3} z^3, \nonumber \\
	&& f_\gamma^{(1)} (z)= \ln z   \left(  - \frac{10}{3} + \frac{2}{z} + 4 z \right)
	+ \ln (1-z)   \left(  - \frac{5}{3} + \frac{1}{z} + 2 z - 2 z^2 + \frac{2}{3} z^3 \right) \nonumber
	\\
	&&    + \frac{1}{3} - \frac{1}{z} + \frac{35}{12} z - 2 z^2 - \frac{1}{4} z^3 .\nonumber
\end{eqnarray}

%\section*{Evolution equation}

Process-independent fragmentation functions were calculated by solving the QED evolution equations
\begin{eqnarray} \label{evol_eq}
	D_{ba}(x, \frac{\mu^2_R}{\mu^2_F}) = \delta(1-x)\delta_{ba} + \sum\limits_{i=e,\bar{e},\gamma}\int\limits_{\mu_R^2}^{\mu_F^2} 
	\frac{dt \alpha(t)}{2 \pi t}  \int\limits_{x}^{1} \frac{dy}{y} D_{ia}(y, \frac{\mu^2_R}{t}) P_{bi} \left( \frac{x}{y}, t \right), 
\end{eqnarray}
by iterations. Details of the calculation of the QED PDFs and fragmentation functions can be found in~\cite{Arbuzov_2023}. %We use the complete expression for running coupling in the 
%$\overline{\mathrm{MS}}$ scheme that can be found in \cite{Baikov:2012rr}.

\section*{Corrections to muon decay spectrum}

The expression for the differential distribution of electrons for unpolarized muon decay, see e.g.~\cite{Arbuzov:2002rp}, reads
\begin{equation}
\frac{d \Gamma}{d z} = \Gamma_0 F(z), \quad \Gamma_0 = \frac{G_F^2 m_\mu^5}{192 \pi^3}, \
 z = \frac{2 m_\mu E_e}{m^2_\mu + m_e^2}, \quad z_0 \leq z \leq 1, \quad z_0 = \frac{2 m_\mu m_e}{m^2_\mu + m_e^2},
\end{equation}
where $G_F$ is the Fermi coupling constant, $E_e$ and $z$ are the electron energy and its energy fraction, and
\begin{equation}
F(z) = f_e^{(0)}(z) + \sum_{i=1,j\leq i}^{\infty}  \alpha^i L^j F_{ij} (z).
\end{equation}
Within the QED PDF approach, a part of higher order corrections to unpolarized muon decay spectrum was calculated in~\cite{Arbuzov:2002rp,Arbuzov:2002cn}. We recalculated the corrections of the orders $\mathcal{O}(\alpha^2 L)$ and $\mathcal{O}(\alpha^3 L^3)$. In $\mathcal{O}(\alpha^2 L)$ we found an additional term $d_{\gamma e}^{(1)}(x) \otimes P_{e \gamma}^{(0)} $ in the fragmentation function $D_{ee}$, which comes from the iterative solution of the evolution equation. In $\mathcal{O}(\alpha^3 L^3)$ we have corrected the coefficient in $D_{ee}$ in front of the term $ P_{\gamma e}^{(0)} \otimes  P_{e \gamma}^{(0)} $, the coefficient should be $\frac{2}{9}$ instead of $- \frac{1}{9}$. The corrected expression for $D_{ee}$ can be found in \cite{Arbuzov_2023}.

To get the next-to-leading contribution of the order $\mathcal{O}(\alpha^3 L^2)$ we have to make convolutions of the fragmentation functions with the functions $f^i_{e}(z)$ and $f^i_{\gamma}(z)$:
\begin{equation} \label{F32conv}
    \left(f_e^{(0)}(z) + \frac{\alpha}{2 \pi}f_e^{(1)} (z)\right) \otimes \bigl[ D_{ee}\bigr]_T + \left(f_\gamma^{(0)} (z)+ \frac{\alpha}{2 \pi}f_\gamma^{(1)} (z) \right) \otimes \bigl[ D_{e \gamma}\bigr]_T,
\end{equation}
and take only the terms proportional to $\alpha^3 L^2$ from the result. Index $T$ denotes here the (time-like) fragmentation function. The results are given below.

\begin{eqnarray}
&& F_{21}(z) = -\frac{4405}{216} + \frac{2 \zeta_2 z^3}{3}-9 \zeta_2 z^2+\left(8 \zeta_2 z^3-12 \zeta_2 z^2-\frac{32 z^3}{9}-19 z^2-13 z-\frac{97}{12}\right) \ln (z) \nonumber \\
&& +12 \zeta_2 z+\left(8 z^3-12 z^2\right) \left(-\mathrm{Li}_3(z)+\mathrm{Li}_2(z) \ln (z)+\frac{1}{2} \ln (1-z) \ln ^2(z)+\zeta_3\right)\nonumber \\
&& +\left(-\frac{16 z^3}{3}+6 z^2-6 z\right) \mathrm{Li}_2(1-z)+\left(24 z^2-16 z^3\right) \mathrm{Li}_3(1-z) \nonumber \\
&& +\left(16 z^3-24 z^2\right)
   \mathrm{Li}_2(1-z) \ln (1-z)+\left(8 z^3-12 z^2\right) \mathrm{Li}_2(1-z) \ln (z)-12 z^3 \zeta_3 \nonumber \\
&& -\frac{167 z^3}{54}+\left(\frac{16 z^3}{3}-12
   z\right) \ln ^2(1-z)+18 z^2 \zeta_3+\frac{449 z^2}{9}+\left(12 z^2-8 z^3\right) \ln ^3(z)\nonumber \\
&& +\left(-\frac{32 z^3}{3}+11 z^2-3 z-\frac{5}{4}\right)
   \ln ^2(z)+\left(24 z^3-36 z^2\right) \ln (1-z) \ln ^2(z)\nonumber \\
&& +\left(12 z^2-8 z^3\right) \ln ^2(1-z) \ln (z)+\left(-\frac{8 z^3}{9}+\frac{4
   z^2}{3}-16 z+\frac{2}{3 z}  -\frac{8}{3}\right) \ln (1-z) \nonumber \\
&&+\left(\frac{8 z^3}{3}-14 z^2+22 z+\frac{20}{3}\right) \ln (1-z) \ln (z)-\frac{1195 z}{36}-\frac{3}{z},
\end{eqnarray}

\begin{eqnarray}
&& F_{32} = \frac{53623}{1296}+\frac{1}{108 z} +\frac{1201 z^3}{162}-\frac{2131 z^2}{72}-\frac{49 z}{2}+\left(8 z^3-4 z^2-12 z\right) \ln ^3(1-z) \nonumber \\
&& +\left(\frac{92 z^3}{9}-\frac{41 z^2}{3}-\frac{7
   z}{3}-\frac{35}{36}\right) \ln ^3(z)+\biggl[\frac{142 z^3}{9}+\frac{152 z^2}{3}+\frac{161 z}{12} +\zeta_2
   \left(60 z^2-40 z^3\right)\nonumber \\
&& +\left(4 z^3-6 z^2\right) \ln ^2(1-z) +\left(-\frac{56 z^3}{3}+58 z^2+44 z+\frac{125}{6}\right) \ln (1-z)+\frac{37}{8} \biggr] \ln ^2(z)\nonumber \\
&& +\left(4 z^3-6
   z^2\right) \mathrm{Li}_2(1-z){}^2 +\zeta_2 \left(-6 z^2+16 z+\frac{139}{18}\right)+\zeta_4 \left(60 z^2-40 z^3\right)\nonumber \\
&& +\zeta_2^2 \left(36
   z^3-54 z^2\right)+\left(104 z^3-156 z^2\right) \mathrm{H}(3,0,z) +\left(80 z^3-120 z^2\right) \mathrm{H}(1,2,0,z)\nonumber \\
&& +\left(104 z^3-156
   z^2\right) \mathrm{H}(2,0,0,z)+\left(104 z^3-156 z^2\right) \mathrm{H}(2,1,0,z) 
\nonumber \\
&& +\left(64 z^3-96 z^2\right)
   \mathrm{H}(0,0,0,0,z)+\left(56 z^3-84 z^2\right) \mathrm{H}(1,0,0,0,z)\nonumber \\
&& +\left(96 z^3-144 z^2\right) \mathrm{H}(1,1,0,0,z) +\left(80
   z^3-120 z^2\right) \mathrm{H}(1,1,1,0,z)\nonumber \\
&&+\left(\frac{136 z^3}{9}+\frac{185 z^2}{3}-\frac{247 z}{3}+\zeta_2 \left(60 z^2-40
   z^3\right)-\frac{160}{3}-\frac{6}{z}\right) \mathrm{Li}_2(1-z) \nonumber \\
&& +\ln ^2(1-z) \biggl(-\frac{62 z^3}{9}+\frac{37 z^2}{6}-\frac{62 z}{3}+\zeta_2
   \left(40 z^3-60 z^2\right)+\left(16 z^3-24 z^2\right) \mathrm{Li}_2(1-z)\nonumber \\
&& -\frac{121}{18}+\frac{1}{z}\biggr) +\left(32 z^3-40 z^2+4 z-\frac{10}{3}\right)
   \mathrm{Li}_3(1-z)+\biggl(\frac{64 z^3}{3}+72 z^2+97 z \nonumber \\
&& +\frac{140}{3}\biggr) \mathrm{Li}_3(z)  +\left(72 z^2-48 z^3\right) \mathrm{Li}_4(1-z)+\left(120
   z^3-180 z^2\right) \mathrm{Li}_4(z) \nonumber \\  
&&    +\left(120 z^2-80 z^3\right) \mathrm{S}_{2,2}(z) +\ln (z) \biggl[ \frac{283 z^3}{27}-\frac{799 z^2}{12}+\frac{539
   z}{36}  +\left(12 z^2-8 z^3\right) \ln ^3(1-z)\nonumber \\
&&  +\left(-\frac{16 z^3}{3}-8 z^2+36 z+10\right) \ln ^2(1-z) +\zeta_2 \left(-\frac{100 z^3}{3}-32
   z^2-125 z-\frac{160}{3}\right) \nonumber \\
&& +\left(\frac{8 z^3}{3}+22 z^2+74 z+\frac{185}{6}\right) \mathrm{Li}_2(1-z) +\ln (1-z) \biggl(-\frac{4 z^3}{3}-11
   z^2-\frac{z}{2}\nonumber \\
&& +\zeta_2 \left(16 z^3-24 z^2\right) +\left(8 z^3-12 z^2\right) \mathrm{Li}_2(1-z) -\frac{57}{4}+\frac{8}{3 z}\biggr) +\left(48 z^2-32
   z^3\right) \zeta_3+ \frac{1261}{108}\biggr]\nonumber \\
&&+\ln (1-z) \biggl[ -\frac{155 z^3}{27}+\frac{2221 z^2}{36}-\frac{677 z}{36} +\zeta_2 \left(-\frac{20
   z^3}{3}-14 z^2+36 z\right)  +\biggl(-32 z^3+ \nonumber \\
&&40 z^2-4 z+\frac{10}{3}\biggr) \mathrm{Li}_2(1-z) +\left(48 z^2-32 z^3\right) \mathrm{Li}_3(1-z)+\biggl(144 
   z^2  -96 z^3\biggr) \mathrm{Li}_3(z) \nonumber \\
&&+\left(88 z^3-132 z^2\right) \zeta_3-\frac{6281}{108}-\frac{32}{9 z} \biggr]+\left(\frac{8 z^3}{3}-92 z^2-109
   z-\frac{125}{3}\right) \zeta_3, \\
&& F_{33} (z) = -\frac{619}{1296} +\frac{20 z}{9}+\frac{4}{27 z}+\frac{16 z^3}{81}-\frac{5 z^2}{9} +\zeta_2 \left(-\frac{16 z^3}{3}+\frac{8 z^2}{3}+\frac{2 z}{3}+\frac{5}{18}\right)\nonumber \\
&&+ \ln (z) \biggl[\zeta_2 \left(12 z^2-8 z^3\right)+\left(\frac{16 z^3}{3}-8 z^2\right) \mathrm{Li}_2(z)+\frac{32 z^3}{27} +\frac{52 z^2}{9} -\frac{67 z}{36}-\frac{41}{108}\nonumber \\
&& +\left(8
   z^3-12 z^2\right) \ln ^2(1-z)+\left(-\frac{8 z^3}{3}+4 z^2-4 z-\frac{5}{3}\right) \ln (1-z)\biggr]  \nonumber \\
&& +\ln (1-z)
   \biggl[\zeta_2 \left(8 z^3-12 z^2\right)+\left(8 z^3-12 z^2\right) \mathrm{Li}_2(1-z)-\frac{32 z^3}{27}-\frac{44 z^2}{9}+\frac{15 z}{2}+\frac{4}{9
   z} \nonumber \\
&& +\frac{289}{108}\biggr] \ +\left(\frac{8 z^3}{3}+\frac{4
   z^2}{3}-\frac{14 z}{3}-\frac{35}{18}\right) \mathrm{Li}_2(z) +\left(12 z^2-8 z^3\right) \mathrm{Li}_3(1-z) \nonumber \\
&&+\left(\frac{8 z^3}{3}-4 z^2+4 z+\frac{5}{3}\right) \ln ^2(1-z) +\biggl(\frac{4 z^3}{3}  -\frac{2 z^2}{3}+\left(2 z^2-\frac{4 z^3}{3}\right) \ln
   (1-z)-\frac{z}{2}-\frac{5}{24}\biggr) \ln ^2(z) \nonumber \\
&&  +\left(8 z^2-\frac{16 z^3}{3}\right)
   \mathrm{Li}_3(z)+\left(4 z^2-\frac{8 z^3}{3}\right) \ln ^3(1-z)+ \left(\frac{4 z^3}{9}-\frac{2 z^2}{3}\right) \ln
   ^3(z) .
\end{eqnarray}

\section*{Conclusions}

So, we presented new results for the QED corrections to unpolarized muon decay spectrum in the $\mathcal{O}(\alpha^3 L^2)$ order. Two corrections to the  known results~\cite{Arbuzov:2002cn, Arbuzov:2002rp} in the $\mathcal{O}(\alpha^2 L)$ and $\mathcal{O}(\alpha^3 L^3)$ orders are found. 
Details of our calculations and comparisons with
the recent result given in~\cite{Banerjee:2022nbr}
will be presented elsewhere~\cite{in_preparation}.
Our results can be useful for new experiments on muon and tau lepton decays, including searches for new physics. 

%% If you have bibdatabase file and want bibtex to generate the
%% bibitems, please use
%%
%\bibliographystyle{elsarticle-num} 
%\bibliography{Mu_decay_corrections}

%% else use the following coding to input the bibitems directly in the
%% TeX file.

\end{document}